\documentstyle[seceq,epsf,wrapfig,twoside]{ptptex}
\notypesetlogo  
\title{Exclusive Weak Radiative Decays of ${\mib B}$ Mesons
in the Covariant Oscillator Quark Model }
\author{%
Rukmani {\sc Mohanta}, Anjan K. {\sc Giri}, Mohinder P. {\sc Khanna}\\
Muneyuki {\sc Ishida}$^{*}$ and Shin {\sc Ishida}$^{**}$
}
\inst{%
Department of Physics,
Panjab University, Chandigarh-160014, India\\
$^{*}$ Department of Physics, Tokyo Institute of Technology\\
Tokyo 152-8551, Japan\\
$^{**}$ Atomic Energy Research Institute, 
College of Science and Technology\\
Nihon University, Tokyo 101, Japan\\ 
}
\recdate{%
December 26, 1998
}
\abst{%
Exclusive weak radiative $B$ meson decays are studied using
the covariant oscillator quark model. The branching ratios
for the processes $B^0 \to K^{*0} \gamma $ and $B^{\pm}
\to K^{* \pm} \gamma $ have been estimated. These are in reasonable
agreement with the available experimental data. The calculation
has been extended to the CKM-suppressed decay processes $B^-
\to \rho^- \gamma $ and $B_s \to K^{*0} \gamma $.
}

\begin{document}
\maketitle

\setcounter{tocdepth}{4}

\section{Introduction}
The study of the weak radiative decay $B \to K^* \gamma $ as a test
of Standard Model (SM) has attracted considerable attention since the
CLEO experiment \cite{ref1} gave the preliminary determination
of the exclusive branching ratio $Br (B \to K^* \gamma )=(4.0
\pm 1.7 \pm 0.8 ) \times 10^{-5} $. The weak radiative decays
of $B$ mesons (which proceed through a flavor changing neutral
current, absent at the tree level in the SM) are remarkable
for several reasons. The $B \to K^* \gamma $ decay arises from
the quark level process $b \to s \gamma $ via penguin-type
diagrams at the one loop level. Hence it is not only a significant test of
Standard Model flavor-changing neutral current dynamics but also
is sensitive to new physics appearing through virtual particles
such as the top quark and $W$ boson in the internal loop. The study
of this process provides valuable information concerning the 
Cabibbo-Kobayashi-Maskawa (CKM) parameters, $V_{td},~V_{ts}$ and $V_{tb}$.
Furthermore, additional contributions in loop stemming from new
bosons and fermions present in most of the extensions of the SM, suggests
the possibility that this process provides a window to new
physics.

The theoretical analysis of the decay process $B \to K^* \gamma $
requires long-distance QCD contributions which cannot
be determined perturbatively. It is also not straightforward to
calculate the exclusive decays by first principles of QCD,
due to complications inherent in nonperturbative
QCD. Therefore, one must resort to some phenomenological models
to obtain reliable results. The heavy quark effective theory
(HQET) \cite{ref2} is expected to be useful in this regard in so
far as the $b$ quark is concerned. However the $s$ quark in the
final hadron can neither be considered heavy enough to enable the
use of HQET nor sufficiently light to permit the exploitation of the
chiral perturbation theory in an unambiguous manner.
There are several methods available in the literature to study the
exclusive $B\to K^* \gamma $ decay process. Some of them include
the QCD sum rule,\cite{ref3,ref31,ref32}
lattice QCD,\cite{ref4} nonrelativistic
and relativistic quark models.\cite{ref5,ref6,ref7,ref8,ref9}
The HQET \cite{ref10} has also been applied to this decay process
even though the $s$-quark mass is certainly not heavy enough, contrary
to the requirement of HQET.

In this paper we study the rare radiative decay
process $B \to K^* \gamma $ using the
covariant oscillator quark model (COQM).\cite{ref11}
One of the most important motives for this
model is to describe covariantly the center of
mass motion of hadrons, while preserving the considerable success of
the non-relativistic quark model regarding the static properties of
hadrons. A keystone in COQM for doing this is treating directly
the squared masses of hadrons in contrast to the mass itself, as done 
in conventional approaches. This makes the covariant treatment
simple. The COQM has been applied to various problems
\cite{ref12} with satisfactory results. Recently Ishida et al.
\cite{ref13,ref181} have studied the weak decays of heavy 
hadrons using this model
and derived the same relations of weak form factors 
for heavy-to-heavy transition as done in HQET.\cite{ref2} 
In addition, our model is also applicable to heavy-to-light
transitions. As a consequence, this model does incorporate
the features of heavy quark symmetry and can be used to
compute the form factors for heavy-to-light transitions as well,
which is beyond the scope of HQET. Actually, in previous
papers we made analyses of the spectra of exclusive 
semi-leptonic\cite{ref181} decays of $B$-mesons
and analyses of non-leptonic decays of $B$ mesons\cite{meson} and 
of hadronic weak decays of 
$\Lambda_b$ baryons\cite{baryon} using this line of
reasoning, leading to encouraging results.
Keeping this
success in mind, we extend its application to
weak radiative decays of $B$ mesons.

The paper is organized as follows. In \S 2 we present
the expressions for the decay widths for the weak radiative
decays of $B$ mesons. In \S 3 we present a brief description
of the covariant oscillator quark model. Using this model
we have evaluated the required form factors.
Section 4 contains our results
and discussion.

\section{Methodology}
The general amplitude of weak radiative decay with one real
photon emission is given by
\begin{eqnarray}
{\cal M}(B(p) &\to & P^*(k) \gamma(q) )=i \epsilon_{\mu
\nu \alpha \beta}~\eta^\mu~q^\nu~ \epsilon^{\alpha}~k^{\beta}~
f_1(q^2) \nonumber\\
& +& \eta^{\mu}\left [\epsilon_{\mu}(M_B^2-M_{P^*}^2)
-(p+k)_{\mu} (\epsilon \cdot q) \right ] f_2(q^2)\;,
\label{eq1}
\end{eqnarray}
where $\eta $ and $\epsilon $ are the polarization vectors of the
photon and the vector meson $P^* $, respectively. The first (second)
term on the right-hand side of Eq. (\ref{eq1}) is a parity conserving
(violating) term, and because of the real photon we have
$q^2$=0. The decay width implied by this
amplitude is given as
\begin{equation}
\Gamma(B \to P^* \gamma)= \frac{1}{32 \pi}\frac{\left (M_B^2
-M_{P^*}^2 \right )^3}{M_B^3} \left [|f_1(0)|^2 +4|f_2(0)|^2
\right ]\;.
\label{eq2}
\end{equation}

Now, the exclusive decay $B \to K^* \gamma $ is expected to be 
described well by the quark level process $b \to s \gamma $ if the 
confinement effect is properly taken into account. 
Accordingly, in the Standard
Model, $B$ decays are described by the effective Hamiltonian obtained
by integrating out the top quark and $W$ boson fields given as
\cite{ref14}
\begin{equation}
{\cal H}_{\rm eff}(b \to s \gamma) = -\frac{4G_F}{\sqrt{2}}\; V_{tb}\;
V_{ts}^*\; \sum_{i=1}^8~C_i(\mu)~O_i(\mu)\;,
\end{equation}
where the $C_i(\mu)$ are the Wilson coefficients,
arise from the renormalization group equation to provide the
scaling down of the subtraction point appropriate to the problem
i.e., $ \mu \approx m_b $. The set $\{O_i\} $ is a complete set of
renormalized, dimension-six operators which govern the $b \to s$
transition. They consist of two current operators $O_1$ and $O_2$
and four strong penguin operators $O_3$--$O_6$, which determine
the nonleptonic decays, and the electromagnetic dipole
operator $O_7 $ and the chromomagnetic dipole operator $O_8$,
which are responsible for rare $B$ decays, i.e., $b \to s+\gamma $
and $b \to s+g $, respectively. Out of the eight operators
$O_i$, the one that contributes to $B \to K^* \gamma $ is
\begin{equation}
O_7=\frac{ e}{32 \pi^2} F_{\mu \nu}~ \left [
m_b \bar s ~\sigma^{\mu \nu}~(1+\gamma_5)~b
+m_s \bar s ~\sigma^{\mu \nu}~(1-\gamma_5)~b \right ]\;,
\end{equation}
where $F_{\mu \nu} $ is the electromagnetic field strength tensor.
The relevant Wilson coefficient is given by \cite{ref14}
\begin{eqnarray}
C_7(\mu) &=& \left [ \frac{\alpha_s(M_W)}{\alpha_s(m_b)} \right
]^{16/23}
\biggr\{C_7(M_W)-\frac{8}{3}C_8(M_W)\nonumber\\
&\times &\left [1-\left (\frac{\alpha_s(m_b)}{\alpha_s(M_W)}
\right )^{2/23} \right ] +\frac{232}{513}
\left [1-\left (\frac{\alpha_s(m_b)}{\alpha_s(M_W)} \right )^{19/23}
\right ] \biggr\}\;,
\end{eqnarray}
with
\begin{equation}
C_7(M_W)=-\frac{x}{2}\left [ \frac{\frac{2}{3}x^2+
\frac{5}{12}x-\frac{7}{12}}{(x-1)^3}-
\frac{(\frac{3}{2}x^2-x){\rm ln}x}{(x-1)^4} \right ]
\end{equation}
and
\begin{equation}
C_8(M_W)=-\frac{x}{4}\left [ \frac{\frac{1}{2}x^2-
\frac{5}{2}x-1}{(x-1)^3}+
\frac{3x{\rm ln}x}{(x-1)^4} \right ]\;,
\end{equation}
where $x=m_t^2/M_W^2 $. From the fact that $m_b \gg m_s $, only
the term involving $m_b $ in the operator $O_7 $  need be
retained. Thus the matrix element of interest becomes
\begin{equation}
\langle K^* \gamma | O_7 | B \rangle =\frac{i e}{16 \pi^2}
~q_{\mu} \eta_{\nu}~m_b~ \langle K^*|\bar s~
\sigma^{\mu \nu}~(1+\gamma_5)~b |B \rangle \;,
\label{eq8}
\end{equation}
where $q_{\mu} $ is the four momentum of the photon and $\eta_{\mu}
$ is its polarization vector. It should be noted that here
the matrix element of a tensor current between hadronic states
for which not much information is available is involved. However,
in the framework of
HQET the associated heavy quark spin symmetry
enables one to express the matrix element of the tensor operator
in terms of the vector and axial vector form factors that also
occur in semileptonic decays and may be estimated in different
phenomenological models.

In the static limit of a heavy $b$ quark we may use the equation
of motion $\gamma_0 b=b $ to derive the relations \cite{ref15}
\begin{equation}
\langle K^*|\bar s~i~\sigma_{0i}~(1+\gamma_5)~b|B \rangle
=\langle K^*|\bar s~\gamma_{i}~(1-\gamma_5)~b|B \rangle \;.
\label{eq:eqn1}
\end{equation}
As a result, the form factors $f_1 $ and $f_2$ in Eq. (\ref{eq1}) can
be related to the vector and axial vector form factors $V$ and $A_1$
appearing in the matrix element of the RHS of Eq. (\ref{eq:eqn1})
defined by \cite{ref16}
\begin{equation}
\langle K^* (k) | \bar s \gamma_{\mu} b |B(p) \rangle
=\frac{2i}{M_B+M_{K^*}} \epsilon_{\mu \nu \alpha \beta}~
\epsilon^{\nu} q^{\alpha}p^{\beta}~ V(q^2)\;,\label{eq:eqn3}
\label{eq10}
\end{equation}

\begin{eqnarray}
\langle K^* (k) | \bar s \gamma_{\mu}\gamma_5 b |B(p) \rangle
&=&(M_B+M_{K^*}) \epsilon_{\mu}~A_1(q^2)-\frac{\epsilon
\cdot q}{M_B+M_{K^*}}(p+k)_{\mu}~A_2(q^2)\nonumber\\
& & -2\frac{\epsilon \cdot q}{q^2} q_{\mu}~ M_{K^*}~[A_3(q^2)
-A_0(q^2)]\;,
\label{eq:eqn4}
\end{eqnarray}
with $q=p-k$. Here $A_3(q^2)$ is simply an abbreviation for
\begin{equation}
A_3(q^2)=\frac{M_B + M_{K^*}}{2 M_{K^*}}~A_1(q^2)
-\frac{M_B-M_{K^*}}{2M_{K^*}}~A_2(q^2)\;,
\label{eq12}
\end{equation}
and in order to cancel the singularity at $q^2=0$,
we must have $A_3(q^2=0)=A_0(q^2=0)$.
Thus with Eqs. (\ref{eq1}) and (\ref{eq8})--(\ref{eq12}),
we obtain the following relations for
$B \to K^* $ transition at $q^2=0 $:\cite{ref17}
\begin{equation}
f_1(0)=\frac{G_F}{\sqrt 2}
\frac{ e}{2 \pi^2}~C_7(\mu)~ V_{tb} V_{ts}^*~ m_b ~F(0)
\label{eq13}
\end{equation}
and
\begin{equation}
f_2(0)=\frac{1}{2}f_1(0),
\label{eq14}
\end{equation}
where
\begin{equation}
F(0)=\frac{M_B -M_{K^*}}{2M_B}V(0)+\frac{M_B+M_{K^*}}{2M_B} A_1(0).
\label{eq15}
\end{equation}
Substituting the above values of $f_1$ and $f_2$ into Eq. (\ref{eq2}),
we obtain the decay width for $B \to K^* \gamma $ as
\begin{eqnarray}
\Gamma(B \to K^* \gamma)&=&\frac{\alpha~G_F^2~m_b^2}
{128\pi^4 M_B^5} ~
|V_{ts} V_{tb}|^2~|C_7(\mu)|^2   (M_B^2 -M_{K^*}^2)^3   \nonumber\\
& & \times   \biggr[
(M_B +M_{K^*})A_1(0)+(M_B-M_{K^*})V(0) \biggr]^2\;.
\label{eq:eqn2}
\end{eqnarray}

Similarly, the decay width for the CKM suppressed FCNC radiative
transition $b \to d \gamma $, which is responsible for $B^-
\to \rho ^- \gamma $ and $B_s \to K^* \gamma $, is obtained
from Eq. (\ref{eq:eqn2}) with $M_B $ and $M_{K^*} $ replaced by
the corresponding initial and final meson masses. The relevant
CKM factor in this case is $|V_{tb} V_{td}|^2 $.

Now to evaluate the form factors $A_1(0) $ and $V(0) $, we use
the covariant oscillator quark model, which is presented in the next
section.

Here it should be noted that the relations (\ref{eq13}) and 
(\ref{eq14}) are also derivable\cite{ref13} directly from Eq. (\ref{eq8})
in COQM, by using the covariant spin wave functions, that is,
the Bargmann-Wigner spinor functions. Accordingly, the expression
of $F(0)$ in terms of a space-time wave function (Eq. (\ref{eq26})
given later) is also derived directly.

\section{ Model framework and the hadronic form factors}

The general treatment of COQM may be called
the ``boosted $LS$-coupling scheme,'' and the wave-functions, being
tensors in $\tilde U(4) \times O(3,1) $-space, reduce to
those in $SU(2)_{\rm spin} \times O(3)_{\rm orbit} $-space in the
nonrelativistic quark model in the hadron rest frame. The
spinor and space-time portion of the wave functions
satisfy separately the respective covariant equations,
the Bargmann-Wigner (BW) equation for the former, and the
covariant oscillator equation for the latter. The form of
the meson wave function has been determined completely
through the analysis of mass spectra.

In COQM the meson states are described by bi-local fields
$\Phi_A^B(x_{1\mu},x_{2\mu}) $, where $x_{1\mu}(x_{2\mu})$ is
the space time coordinate of the constituent quark (antiquark),
$A=(a,\alpha)\;(B=(b,\beta))$, describing its flavor and
covariant spinor. Here we write only the positive frequency
part of the relevant ground state fields:
\begin{equation}
\Phi_A^B(x_{1\mu},x_{2\mu})=e^{iP\cdot X}\; U(P)_A^B\;
f_{ab}(x_{\mu};P)\;,
\end{equation}
where $U$ and $f$ are the covariant spinor and internal
space-time wave functions, respectively, satisfying the
Bargmann-Wigner and oscillator wave equations. The quantity
$x_{\mu} ~(X_{\mu}) $ is the relative (CM) coordinate,
$x_{\mu}\equiv x_{1\mu}-x_{2\mu} ~(X_{\mu}\equiv m_1x_{1\mu}
+m_2x_{2\mu})/(m_1+m_2) $, with $m_i$ the quark
masses). The function $U$ is given by

\begin{equation}
U(P)=\frac{1}{2 \sqrt 2}\left [(-\gamma_5 P_s(v)+i\gamma_{\mu}
V_{\mu}(v))(1+iv \cdot \gamma)\right ],
\end{equation}
where $P_s(V_\mu )$ represents the pseudoscalar (vector) meson
field, and $v_{\mu} \equiv P_{\mu}/M~ (P_{\mu}(M)$ is the
four momentum (mass) of the meson). The function $U$, being
represented by the direct product of a quark and antiquark
Dirac spinor with the meson velocity, is reduced to the
non-relativistic Pauli-spin function in the meson rest frame.
The function $f$ is given by
\begin{equation}
f(x_{\mu};P)=\frac{\beta}{\pi} \exp\left (-\frac{\beta}{2}
\left (x_{\mu}^2+2\frac{(x\cdot P)^2}{M^2}\right ) \right )\;.
\end{equation}
The value of the parameter $\beta $ is determined from the mass
spectra \cite{ref18} as $\beta_{\rho}=0.14 $, $\beta_{K^*}$
= 0.142, $\beta_B$
= 0.151 and $\beta_{B_s} $= 0.160 (in units of
$ \mbox{GeV}^2$ ).

The effective action for weak interactions of mesons
with $W$-bosons is given by

\begin{equation}
S_W=\int d^4 x_1 d^4 x_2 \langle \bar \Phi_{F, P^\prime}
(x_1,x_2)i \gamma_{\mu}(1+\gamma_5) \Phi_{I, P}(x_1,x_2)
\rangle W_{\mu,q}(x_1)\;,
\label{eq:eqn6}
\end{equation}
where we have denoted the interacting (spectator) quarks as
1 (2) (the KM matrix elements and the coupling constants
are omitted). This is obtained from consideration of
Lorentz covariance, assuming a quark current with the
standard $V-A$ form. In Eq. (\ref{eq:eqn6}), $\Phi_{I,P}~
(\bar \Phi_{F,P^\prime})$ denotes the initial (final) meson
with definite four momentum $P_\mu~(P_\mu^\prime) $, and
$q_{\mu} $ is the momentum of $W$-boson. The function
$\bar \Phi $ is the Pauli-conjugate of $\Phi $ defined by
$\bar \Phi \equiv -\gamma_4 \Phi^\dagger \gamma_4 $, and
$\langle ~~~\rangle$ represents the trace of Dirac spinor
indices. Our relevant effective currents $J_\mu(X)_{P^\prime,P}$
are obtained\cite{ref13} by identifying the above action with

\begin{equation}
S_W =\int d^4 X J_{\mu}(X)_{P^\prime,P}\;W_{\mu}(X)_q,
\end{equation}

\begin{eqnarray}
J_{\mu}&=&I^{qb}(w) \sqrt{M M^\prime}
\times [\bar P_s(v^\prime)P_s(v)(v+v^\prime )_{\mu}\nonumber\\
& & +\bar V_{\lambda}(v^\prime ) P_s(v) (\epsilon_{\mu \lambda
\alpha \beta} ~v_{\alpha}^{\prime}~v_\beta-\delta_{\lambda \mu}
(w+1) - v_{\lambda}~v_{\mu}^{\prime}]\;,
\label{eq:eqn5}
\end{eqnarray}
where $M (M^\prime) $ denotes the physical mass of the
initial (final) meson and $w=-v\cdot v'$. The quantity $I^{qb}(w) $,
the overlapping of the initial
and final wave functions, represents the universal form
factor function.\footnote{
In this paper we apply the pure-confining approximation,
neglecting the effect of the one-gluon-exchange potential.
This approximation is expected to be good for the heavy/light-quark
meson systems.
} It describes the confinement effects
of quarks, and is given by

\begin{equation}
I^{qb}(w)=\frac{4 \beta \beta^\prime}{\beta+\beta^\prime }
\frac{1}{\sqrt{C(w)}} \exp(-G(w))\;;~~
C(w)=(\beta-\beta^{\prime})^2+4\beta \beta^{\prime} w^2\;,
\label{eq23}
\end{equation}
and
\begin{eqnarray}
G(w)&=& \frac{m_n^2}{2 C(w)} \biggr[ (\beta+\beta^\prime)
\left \{ \left (\frac{M}{M_s} \right )^2 + \left (\frac{M^\prime}
{M_s^\prime}\right )^2 -2 \frac{M M^\prime}{M_s M_s^\prime}w
\right \}\nonumber\\
& &+ 2 \left \{ \beta^\prime\left (\frac{M}{M_s} \right )^2
+\beta\left ( \frac{M^\prime}{M_s^\prime}\right )^2
\right \}(w^2-1) \biggr]\;,
\label{eq24}
\end{eqnarray}
where $M_s(M_s^\prime)$ represents the sum of the constituent
quark masses of the initial (final) meson, and $m_n $ is the
spectator quark mass.

Comparing our effective current (\ref{eq:eqn5}) with Eqs.
(\ref{eq:eqn3}) and (\ref{eq:eqn4}), we obtain\footnote{
Here, note the remark given at the end of \S 2.
} the relations
between the invariant form factors $V$ and $A_1$ with the
form factor function $I(w)$  for $ B \to K^* $ transitions
as \cite{ref181}
\begin{eqnarray}
V(0) &=& A_1(0)=F(0), \label{eq25}\\
F(0) &=& \frac{M_B+M_{K^*}}{2 \sqrt{M_B M_{K^*}}} 
I^{sb}(w|_{q^2=0})\;,
\label{eq26}
\end{eqnarray}
and similarly for $B^- \to \rho^- $ and $B_s \to K^* $ transitions
with $I^{db}(w) $ replacing $I^{sb}(w) $.

\section{Results and conclusion}

To estimate the branching ratios for weak radiative decays
of $B$ mesons, we use the
following values for various quantities. The quark masses
(in GeV) are taken as $m_u=m_d$ = 0.4, $m_s$ = 0.55
and $m_b$ = 5. The particle masses and their lifetimes
are taken from Ref. \citen{ref19}. The relevant CKM parameters
required for all the processes considered here are
taken from Ref. \citen{ref19} as $V_{td}=0.0085$, $V_{ts}=
0.0385$ and $V_{tb}=0.99925 $.
The renormalized Wilson coefficient $C_7(m_b) $ used in the
estimation of branching ratios is taken to be
$|C_7(m_b)|$=0.311477.\cite{ref20}
With the above values we first evaluate the form factor
$F(0) $ using Eqs. (\ref{eq15}) and (\ref{eq23})--(\ref{eq25}). 
Our results for
the form factor values given in Table I compare well
with the recent calculations within the framework of the 
relativistic quark model,\cite{ref8} light cone QCD
sum rule,\cite{ref31} and hybrid sum rule.\cite{ref32}
With the calculated values of the form factors, the branching
ratios for different channels are
estimated using Eq. (\ref{eq:eqn2}) as well as the mean
life values of the appropriate decaying meson, which
are given in Table II. The branching ratios $Br(B^0 \to
K^{*0} \gamma $ and $Br (B^{\pm} \to K^{*\pm} \gamma )$ are found to
be in very good agreement with the available data. 
Here it may be worthwhile to note that the relevant process
is relativistic, and the form factor function plays a significant
role, such that $I=0.235$.
Since there are
no data as yet available for the branching ratios in the case
$B \to \rho \gamma $ and $B_s \to K^* \gamma $, the model
predictions must be compared with other theoretical
predictions available in the literature. In fact there are
few theoretical attempts made so far in this sector.
Recently, Singer \cite{sing94} predicted that in the
relativistic quark model,
\begin{equation}
Br(B_s \to K^* \gamma )=(1.4\pm 0.6) \times 10^{-6},
\end{equation}
and in the relativistic independent quark model,
Barik et al.\cite{ref9} obtained
\begin{eqnarray}
&&Br(B \to \rho \gamma)=1.24 \times 10^{-6},\nonumber\\
&&Br(B_s \to K^* \gamma)=9.28 \times 10^{-7}\;.
\end{eqnarray}

Finally, the hadronization ratio $R $ defined as $\Gamma(B \to
K^* \gamma)/\Gamma(b \to s \gamma) $, where
\begin{equation}
\Gamma(b \to s \gamma) =\frac{G_F^2 \alpha}{32 \pi^4}~
~|V_{ts} V_{tb}|^2~|C_7(m_b)|^2~ m_b^5\;,
\end{equation}
is found to be 0.12, which agrees well with the experimentally
observed value $R=0.19 \pm 0.09.$ \cite{ref21}

In view of the consistency of our predictions,
obtained with no free parameters, with the large
number of theoretical predictions as well as the experimental data,
it may be concluded that 
the framework of COQM  provides a suitable scheme 
to estimate the confined effect of quarks, and 
so far as the relevant flavor changing 
neutral current is concerned,  the Standard Model is valid.

\begin{table}
\caption{ Prediction for the rare radiative decay form
factor $F(0) $ in comparison with various theoretical
predictions.}
\vspace {0.2 true in}
\begin{tabular}{l|l|l|l|l}
\hline
\hline
\multicolumn{1}{l|}{Decay process}&
\multicolumn{1}{|l|}{Present work} &
\multicolumn{1}{|l|}{Ref. \citen{ref8}}&
\multicolumn{1}{|l|}{Ref. \citen{ref31}} &
\multicolumn{1}{|l}{Ref. \citen{ref32}}\\
\hline
& & & & \\
$B^{\pm} \rightarrow K^{* \pm} \gamma $ &
0.328 &- &- &-  \\
& & & & \\
$B^{0} \rightarrow K^{*0} \gamma$ &
0.329 & $0.32 \pm 0.03$& $0.32 \pm 0.05 $
&$0.308 \pm 0.013 \pm 0.036 \pm 0.006  $\\
& & & & \\
$B^- \rightarrow \rho^{-} \gamma$ &
0.295 & $ 0.26 \pm 0.03 $ & $ 0.24 \pm 0.04 $ &
$0.27 \pm 0.011 \pm 0.032$\\
& & & & \\
$B_s \to K^{* 0} \gamma $ & 0.250 &
$0.23\pm 0.02 $ & $ 0. 20 \pm 0.04 $& -\\
&&& & \\
\hline
\end{tabular}
\end{table}

\begin{table}
\caption{ Prediction for the branching ratios
of the exclusive rare decays of $B$ mesons
along with their experimental values.}
\vspace {0.2 true in}
\begin{tabular}{l|l|l}
\hline
\hline
\multicolumn{1}{l|}{Decay process}&
\multicolumn{1}{|l|}{Br ratio} &
\multicolumn{1}{|l}{Br ratio}\\
\multicolumn{1}{l|}{} &
\multicolumn{1}{|l|}{This work}&
\multicolumn{1}{|l}{Expt. \citen{ref19}}\\
\hline
 & & \\
$B^0 \rightarrow K^{*0} \gamma $ &
3.96 $\times 10^{-5}$ & $ (4.0 \pm 1.9)
\times 10^{-5}$ \\
& & \\
$B^{\pm} \rightarrow K^{*\pm} \gamma$ &
4.168 $\times 10^{-5}$ & $(5.7 \pm 3.3)\times 10^{-5}$ \\
& & \\
$B^- \rightarrow \rho^{-} \gamma$ &
$1.68 \times 10^{-6}$ &- \\
& & \\
$B_s \to K^{* 0} \gamma $ & $1.158 \times 10^{-6} $ & -\\
& & \\
\hline
\end{tabular}
\end{table}

\newpage

\acknowledgements

R. M. would like to thank CSIR, Govt. of India, for a fellowship.
A. K. G. and M. P. K. acknowledge financial support from
DST, Govt. of India.

\end{document}